\documentclass[12pt,twoside,dvips]{article}
\usepackage{verbatim}
\usepackage{epsfig}
\usepackage{epstopdf}
\input{epsf.sty}
\usepackage{epsf}
\usepackage{amssymb}
\usepackage{amsfonts}
\usepackage{amsmath}
\usepackage{pifont}
\usepackage[spanish,english]{babel}
\usepackage{graphicx}

\begin{document}

\title{Some phenomenological aspects of a new $U(1)'$ model}
\author{R. Mart\'{\i}nez$\thanks{%
e-mail: remartinezm@unal.edu.co}$, J. Nisperuza, F. Ochoa$\thanks{%
e-mail: faochoap@unal.edu.co}$, J. P. Rubio \and Departamento de F\'{\i}sica, Universidad
Nacional de Colombia, \\ Ciudad Universitaria, Bogot\'{a} D.C.}
 
\maketitle

\begin{abstract}
We propose a new non-universal $U(1)'$ extension of the standard model with the addition of three exotic quark singlets, two scalar singlets and one additional scalar doublet. We obtain family dependent couplings with a new $Z'$ boson in the quark sector and universal couplings with the lepton sector. From experimental data on $Z'$ research at CERN-LHC collider, we find limit regions in the $U(1)'$ free parameters ($Z'$ mass and coupling constant), which we use to obtain total decay width and invariant-mass distributions. By introducing discrete symmetries and mixing couplings between ordinary and exotic fermions, we obtain predictable mass relations in the quark sector compatible with the phenomenological values without large fine tuning of the Yukawa couplings and with few free parameters, where hierarchies between quark families can be understood from the existence of heavy beyond standard model particles. 
\end{abstract}

\section{Introduction}

Models with extra $U(1)'$ symmetry is one of the most studied extensions of the Standard Model (SM) \cite{SM}. There are many motivations to consider this kind of models. For example, many grand unified and superstring models contains one or multiple extra $U(1)'$ symmetries in the effective low energy limit \cite{strings}. In supersymmetric extensions, an additional $U(1)'$ factor may suppress the $\mu$-term at tree-level, and provides a mechanisms to generate an effective $\mu _{eff}$ mass through the addition of an scalar singlet \cite{MSSM}.  Non-supersymmetric extensions gives rise to a variety of models \cite{LR, little} (models with dynamical symmetry breaking, little Higgs, extra dimension, left-right models, etc),  which involves a wide number of phenomenological and theoretical aspects, including flavor physics \cite{flavor}, neutrino physics \cite{neutrino}, dark matter \cite{DM}, among other effects. A complete review about the above possibilities can be found in reference \cite{review}.

In particular, family non-universal $U(1)'$ symmetry models have many well-established motivations. For example, they provide hints to solve the SM flavor puzzle, where regardless that all the fermions acquire masses at the same scale $\nu =246$ GeV, experimentally they exhibit very different mass values. These models also implies a new $Z'$ neutral boson, which contains a large number of phenomenological consequences at low and high energies \cite{zprime-review}. Limits on $Z'$ resonances have recently been published at CERN-LHC collider by the ATLAS and CMS collaborations, which push the lower mass bounds to the order of $2$ TeV. In addition of a new neutral gauge boson $Z'$, an extended fermion spectrum is necessary in order to obtain an anomaly-free theory. Also, the new symmetry requires an extended scalar sector in order to {\it i.)} generate the breaking of the new abelian symmetry and {\it ii.)} obtain heavy masses for the new $Z'$ gauge boson and the extra fermion content.

On the other hand, these extensions usually possess specialized Two Higgs Doublet Models in the low energy limit, where two scalar doublets $\phi _1$ and $\phi _2$  are introduced in order to generate the appropriate Yukawa couplings that provide masses to all fermions. However, these type of models predict huge flavor changing neutral currents (FCNC) and CP-violating effects, which are severely suppressed by experimental data at electroweak scales. One way to remove these effects, is by imposing discrete symmetries, which restrict the Yukawa terms and produce additional effects in the hierarchical structures of the fermion masses. 

In this paper, we construct an anomaly-free and family non-universal $U(1)'$ symmetry model with two scalar doublets and the addition of three exotic quarks and two scalar singlets, one of them being candidate to dark matter. In section 2, we show some properties of the particle content of the model. In section 3, we construct the Higgs potential, the Dirac and the Yukawa Lagrangians. In particular, we obtain the mass spectrum of the scalar and the neutral gauge sector, including $Z-Z'$ mixing terms. Section 4 is devoted to study some couplings of the new $Z'$ gauge boson and some phenomenological studies at the LHC Collider. In particular, we perform a $Z'$ production analysis on $pp \rightarrow Z' \rightarrow f\overline{f}$ dispersions, where \textit{i.)} limits on the $Z'$ mass and the $U(1)'$ coupling constant are obtained from experimental limits found by the ATLAS collaboration in dilepton final states, and \textit{ii.)} invariant-mass distributions in $t\overline{t}$ dispersion is obtained. We also construct the Yukawa Lagrangian, where zero-texture mass matrices for the quark sector is obtained by using  an specific $Z_2 \times U(1)_{T_3}$ global symmetry. We found that three quarks (up (u), down (d) and strange (s)) acquire masses at the MeV scale, and three quarks (charm (c), bottom (b) and top (t)) exhibits masses at the GeV scale. Finally, in Sec. 5 we summarize our results.  

\section{The Particle Content}

The proposed model belongs to the class of models with one extra non-universal family $U(1)'$ symmetry which we label as $U(1)_X$. The particle content is composed by the ordinary SM particles and new exotic non-SM particles, as shown in Tabs. \ref{tab:SM-espectro} and \ref{tab:exotic-espectro}, respectively, where column $G_{sm}$ indicates the transformation rules under the SM gauge group $(SU(3)_c,SU(2)_L,U(1)_Y)$, the column $U(1)_X$ are the values of the new quantum number $X$, and in the column labeled as {\it Feature} we describe the type of field. This spectrum exhibits the following properties:   

\begin{itemize}

\item[-] The $U(1)_X$ symmetry is non-universal in the left-handed SM quark sector: the quark family $i=1$ have $X_1=1/3$ while families $i=2,3$ have $X_{2,3}=0$. However, the corresponding right-handed singlets are universal. In this work, we use the normal assignation, where each family $q^i$ correspond to the phenomenological family, i.e.:

\begin{equation}
U^{1,2,3}=(u,c,t), \hspace{1cm} D^{1,2,3}=(d,s,b). 
\label{quark-assignation}
\end{equation}

\item[-]  The SM leptons are family universal but with nontrivial charges $X$.

\item[-]  The scalar doublet $\phi _1$ also has a nontrivial charge $X$.

\item[-] The three extra singlets $T$ and $J^{n}$ are new up-  and down-like quarks, respectively, where $n=1,2$. They are quasi-chiral, i.e. chiral under  $U(1)_X$ and vector-like under $G_{sm}$ 

\item[-] We include new 
neutrinos  $(\nu ^i_R)^c$ and $N ^i_R$ which may generate see-saw neutrino masses in order to obtain a realistic model compatible with oscillation data.

\item[-] The spectrum includes an additional scalar doublet $\phi _2$ identical to $\phi _1$ under  $G_{sm}$ but with different $U(1)_X$ charges, 
where the electroweak scale is related to the VEVs by $\nu = \sqrt{\nu  _1 ^2+\nu _2^2}$. 


\item[-] An extra scalar singlet $\chi _0$ with VEV $\nu _{\chi}$ is required to produce the symmetry breaking of the $U(1)_X$ symmetry. We assume that it happens at a large scale $\nu _{\chi} \gg \nu$.

\item[-] Another scalar singlet $\sigma _0$ is introduced. Since it is not essential for the symmetry breaking mechanisms, we may choose a small VEV $\langle \sigma _0 \rangle = \nu _{\sigma} \lesssim \nu$, Thus, this singlet can work as dark matter candidate. 

\item[-] The extra neutral gauge boson $B'_{\mu}$ is required to obtain a local $U(1)_X$ symmetry.

\item[-] We define the weak hypercharge $Y_f$ as usual, where the electric charge follows the Gell-Mann-Nishijima relation:

\begin{equation}
Q_f=I_3+\frac{Y_f}{2}
\end{equation} 

with $I_3$ the isospin defined for left- and right-handed fermions, according to Eq. (\ref{isospin}).
\end{itemize}

\begin{itemize}

\item[-] A fundamental condition of the model is the cancelation of chiral anomalies. Since the new symmetry introduces an additional gauge boson, there arise new couplings that induce the following nontrivial triangle anomalies:

\begin{eqnarray}
\left[ SU(3)_{c}\right] ^{2} U(1)_{X} &\rightarrow &A_{1}=
\sum _{Q} X_{Q_{L}}-\sum_{Q}X_{Q_R}  \notag 
\\
\left[ SU(2)_{L}\right] ^{2} U(1)_{X} &\rightarrow&A_{2}=
\sum _{\ell} X_{\ell _{L}}+3\sum_{Q}X_{Q_L},  \notag 
\\
\left[ U(1)_{Y}\right] ^{2} U(1)_{X} &\rightarrow&A_{3}=
\sum _{\ell ,Q} \left[Y_{\ell _L}^2X_{\ell _{L}}+3Y_{Q_L}^2X_{Q_{L}}\right]-\sum _{\ell ,Q} \left[Y_{\ell _R}^2X_{\ell _{R}}+3Y_{Q_R}^2X_{Q_{R}}\right]  \notag 
\\
U(1)_{Y} \left[ U(1)_{X} \right] ^{2}&\rightarrow&A_{4}=
\sum _{\ell ,Q} \left[Y_{\ell _L}X_{\ell _{L}}^2+3Y_{Q_L}X_{Q_{L}}^2\right]-\sum _{\ell ,Q} \left[Y_{\ell _R}X_{\ell _{R}}^2+3Y_{Q_R}X_{Q_{R}}^2\right]  \notag 
\\
\left[ U(1)_{X}\right] ^{3} &\rightarrow &A_{5}=
\sum _{\ell ,Q} \left[X_{\ell _{L}}^3+3X_{Q_{L}}^3\right]-\sum _{\ell ,Q} \left[X_{\ell _{R}}^3+3X_{Q_{R}}^3\right] \notag
\\
\left[ Grav\right] ^{2}\otimes U(1)_{X} &\rightarrow&A_{6}=
\sum _{\ell ,Q} \left[X_{\ell _{L}}+3X_{Q_{L}}\right]-\sum _{\ell ,Q} \left[X_{\ell _{R}}+3X_{Q_{R}}\right] 
\label{anomalias}
\end{eqnarray}
where the sums in $Q$ run over all the quarks ($u^i, d^i, T, J^n$), while $\ell $ runs over all leptons with nontrivial $U(1)_X$ values (i.e. $ e^i, \nu _L ^i, (\nu _R^i)^c$). It is a matter of arithmetic to show that the $U(1)_X$ values given in Tabs. \ref{tab:SM-espectro} and \ref{tab:exotic-espectro} are possible solutions that cancel the above anomaly equations.

\end{itemize} 
 
\section{The Lagrangians}

Taking into account  the above particle content, we may construct the complete Lagrangian of the model. In particular, we show the Higgs potential, the mass spectrum of the neutral gauge sector, and the Dirac and Yukawa Lagrangians. 

\subsection{The Higgs Potential}

The most general, renormalizable and $G_{sm} \times U(1)_X$ invariant potential is

\begin{eqnarray}
V&=&\mu _1^2 \phi _1^{\dagger} \phi _1+\mu _2^2 \phi _2^{\dagger} \phi _2 + \mu _3^2 \chi _0^* \chi _0 +\mu _4^2 \sigma _0^* \sigma _0  \nonumber \\
&+&f_1\left(\phi _1^{\dagger} \phi _2 \sigma _0+h.c.\right)+f_2\left(\phi _1^{\dagger} \phi _2 \chi _0+h.c.\right) \nonumber \\
&+& \lambda _1 \left( \phi _1^{\dagger} \phi _1 \right)^2+\lambda _2 \left( \phi _2^{\dagger} \phi _2 \right)^2+\lambda _3 \left(\chi _0^* \chi _0\right)^2+\lambda _4 \left(\sigma _0^* \sigma _0\right)^2 \nonumber \\
&+& \lambda _5 \left( \phi _1^{\dagger} \phi _1 \right)\left( \phi _2^{\dagger} \phi _2 \right)+\lambda '_5 \left( \phi _1^{\dagger} \phi _2 \right)\left( \phi _2^{\dagger} \phi _1 \right)+ \lambda _6 \left( \phi _1^{\dagger} \phi _1 \right)\left(\chi _0^* \chi _0\right)\nonumber \\
&+&\lambda '_6 \left( \phi _1^{\dagger} \phi _1 \right)\left(\sigma _0^* \sigma _0\right)+ \lambda _7 \left( \phi _2^{\dagger} \phi _2 \right)\left(\chi _0^* \chi _0\right)+\lambda '_7 \left( \phi _2^{\dagger} \phi _2 \right)\left(\sigma _0^* \sigma _0\right) \nonumber \\
&+&\lambda _8  \left(\chi _0^* \chi _0\right)\left(\sigma _0^* \sigma _0\right)+\lambda '_8  \left[\left(\chi _0^* \sigma _0\right)\left(\chi _0^* \sigma _0\right)+h.c.\right].
\label{higgs-pot}
\end{eqnarray}

When we apply the minimum conditions $\partial \langle V \rangle/\partial \nu _i$ for each scalar VEV $\nu _i = \nu _{1,2,\chi, \sigma}$, the following relations are obtained:

\begin{eqnarray}
\mu _1^2&=&-\frac{1}{\sqrt{2}}(f_1\nu _{\sigma}+f_2\nu _{\chi})\frac{\nu _2}{\nu _1}-\lambda _1 \nu _1^2-\frac{1}{2}(\lambda _5+\lambda '_5)\nu _2^2-\frac{1}{2}\lambda _6\nu _{\chi }^2-\frac{1}{2}\lambda '_6\nu _{\sigma }^2, \nonumber \\
\mu _2^2&=&-\frac{1}{\sqrt{2}}(f_1\nu _{\sigma}+f_2\nu _{\chi})\frac{\nu _1}{\nu _2}-\lambda _2 \nu _2^2-\frac{1}{2}(\lambda _5+\lambda '_5)\nu _1^2-\frac{1}{2}\lambda _7\nu _{\chi }^2-\frac{1}{2}\lambda '_7\nu _{\sigma }^2, \nonumber \\
\mu _3^2&=&-\frac{f_2}{\sqrt{2}}\frac{\nu _1\nu _2}{\nu _{\chi }}-\lambda _3 \nu _{\chi }^2-\frac{1}{2}\lambda _6\nu _1^2-\frac{1}{2}\lambda _{7 }\nu _2^2-\frac{1}{2}(\lambda _8+2\lambda' _8)\nu _{\sigma} ^2, \nonumber \\
\mu _4^2&=&-\frac{f_1}{\sqrt{2}}\frac{\nu _1\nu _2}{\nu _{\sigma }}-\lambda _4 \nu _{\sigma }^2-\frac{1}{2}\lambda '_6\nu _1^2-\frac{1}{2}\lambda '_{7 }\nu _2^2-\frac{1}{2}(\lambda _8+2\lambda' _8)\nu _{\chi} ^2.
\end{eqnarray}

With the above parameters replaced in (\ref{higgs-pot}), we find the square mass matrices $M_{R}^2$ for the real fields, $M_{I}^2$ for the imaginary fields and $M_{C}^2$ for the charged fields. For the simplest case with $\nu _{\sigma}=0$, we obtain:

\begin{eqnarray}
M_R^2&=&\begin{pmatrix}
8\lambda _1\nu _1^2-2f_2\frac{\nu _2\nu _{\chi}}{\nu _{1}} & 2f_2\nu _{\chi}+4(\lambda _5+\lambda '_5)\nu _1\nu _2 & 2f_2\nu _2+4\lambda _6\nu _1\nu _{\chi}
 \\
* & 8\lambda _2\nu _2^2-2f_2\frac{\nu _1\nu _{\chi}}{\nu _{2}}  &  2f_2\nu _1+4\lambda _7\nu _2\nu _{\chi} \\
* & * &  8\lambda _3\nu _{\chi }^2-2f_2\frac{\nu _1\nu _{2}}{\nu _{\chi }}  \\
\end{pmatrix},
\label{real-scalar-mass}
\end{eqnarray}
in the basis ${\xi _1,\xi _2, \xi _{\chi}}$,

\begin{eqnarray}
M_I^2&=&\begin{pmatrix}
-2f_2\frac{\nu _1\nu _{\chi}}{\nu _{2}} & 2f_2\nu _{\chi} & 2f_2\nu _1
 \\
* & -2f_2\frac{\nu _2\nu _{\chi}}{\nu _{1}}  &  -2f_2\nu _2 \\
* & * &  -2f_2\frac{\nu _1\nu _{2}}{\nu _{\chi }}  \\
\end{pmatrix}, 
\label{imag-scalar-mass}
\end{eqnarray}
in the basis ${\phi _2^0,\phi _1^0, \zeta _{\chi}}$, and

\begin{eqnarray}
M_C^2&=&\begin{pmatrix}
-f_2\frac{\nu _1\nu _{\chi}}{\nu _{2}}-\lambda '_5\nu _1^2 & f_2\nu _{\chi}+\lambda '_5\nu _1\nu _2
 \\
* & -f_2\frac{\nu _2\nu _{\chi}}{\nu _{1}}-\lambda '_5\nu _2^2  \\
\end{pmatrix},
\label{charge-scalar-mass} 
\end{eqnarray}
in the basis ${\phi _2^+,\phi _1^-}$. To obtain the eigenvalues and eigenvectors, we assume the hierarchy $f_2\nu _{\chi} \gg \nu _{1,2}^2$. After diagonalization, we obtain the following physical spectrum and their squared masses:

\begin{eqnarray}
m_{h_0}^2 &\approx& \nu ^2\left[\lambda _2 S_{\beta}^4 +\lambda _1C_{\beta}^4+(\lambda _5+\lambda '_5)C_{\beta}^2S_{\beta }^2\right] \nonumber \\
m_{H_0}^2 &\approx& \frac{2f_2\nu _{\chi}}{C_{\beta}S_{\beta}}, \nonumber \\
m_{H_\chi ^0}^2&\approx&8\lambda _3\nu_{\chi}^2
\end{eqnarray}
for the real sector,

\begin{eqnarray}
m_{A_{0}}^2&=&\frac{m_{H_0}^2}{2}\left[1+C_{\beta}^2S_{\beta}^2\left(\frac{\nu }{\nu _{\chi}}\right)^2\right]
\end{eqnarray}
corresponding to a pseudoscalar boson, and

\begin{eqnarray}
m_{H^{\pm}}^2 = \frac{m_{H_0}^2}{2}\left[1+\lambda '_5C_{\beta}S_{\beta}\left(\frac{\nu ^2}{2f_2\nu _{\chi}}\right) \right] 
\end{eqnarray}
a charged Higgs boson. In the above expresions, we define the electroweak VEV as $\nu = \sqrt{\nu _1^2+\nu _2^2}$, and the angle

\begin{equation}
\tan(\beta)=\frac{\nu _1}{\nu _{2}}.
\label{beta-angle}
\end{equation}

In addition, we obtain two charged and two neutral Goldstone bosons, which will give masses to two charged ($W^{\pm}$) and two neutral ($Z$ and $Z'$) gauge bosons, respectively.
 
\subsection{Neutral gauge masses}

The kinetic sector of the Higgs Lagrangian reads:

\begin{eqnarray}
\mathcal{L}_{kin} &=&\sum_i (D_{\mu}\Phi _i)^{\dagger}(D^{\mu}\Phi _i),
\label{higgs-kinetic}
\end{eqnarray}
where $\Phi _i=\phi _{1,2},\chi _0, \sigma _0$, and the covariant derivative is defined as

\begin{eqnarray}
D_{\mu}=\partial _{\mu }-igW_{\mu}^{\alpha} T_{\alpha}-ig'\frac{Y}{2}B_{\mu}-ig_XXB'_{\mu}.
\label{covariant}
\end{eqnarray}
After expanding the terms evaluated in the VEVs of the scalar fields, and defining $\nu =\sqrt{\nu _1^2+\nu _2^2 }$ the electroweak VEV, $\nu _S=\sqrt{\nu _{\sigma}^2+\nu _{\chi}^2 }$ the singlet VEV, and  $\epsilon =\nu /\nu _S$, we obtain the following symmetric mass matrix in the neutral gauge basis  $({W_{\mu }^3,B_{\mu},B'_{\mu}})$:

\begin{eqnarray}
M_0^2&=&\frac{1}{4}\begin{pmatrix}
g^2\nu ^2 &-gg'\nu ^2   &-\frac{2}{3}gg_X\nu ^2(1+S_{\beta} ^2)   \\ 
&&\\
* &  g'^2\nu ^2 &  \frac{2}{3}g'g_X\nu ^2(1+S_{\beta}^2)   \\
 &&\\
* &*  &  \frac{4}{9}g_X^2\nu _S^2\left[1+(1+3S_{\beta}^2)\epsilon^2\right] \\
\end{pmatrix},
\end{eqnarray}
As a first approximation, for $\nu << \nu _S$, the terms proportional to $\epsilon ^2$ are negligible, obtaining the following rotations:

\begin{eqnarray}
A_{\mu }&=&S_WW_{\mu }^3+C_WB_{\mu}, \nonumber \\
Z_{\mu}&=&C_WW_{\mu }^3-S_WB_{\mu}, \nonumber \\
Z'_{\mu}&=&B'_{\mu },
\label{1-eigeneutral}
\end{eqnarray}
with eigenvalues

\begin{eqnarray}
m_A^2&=&0 ,\nonumber \\
m_Z^2&\approx& \frac{g^2\nu ^2}{4C_W^2} ,\nonumber \\
m_{Z'}^2&\approx&  \frac{g_X^2\nu _S^2}{9}.
\end{eqnarray}

Since the current experimental data exclude a $Z'$ boson at the TeV scales, a natural hierarchical relation $\nu _S/\nu  \sim 10$ arise. However, if the above rotations are replaced into the original Lagrangian in (\ref{higgs-kinetic}), we still obtain a $Z-Z'$ mixing term, given by the following mixing matrix in the $(Z,Z')_{\mu}$ basis:

\begin{eqnarray}
M_{ZZ'}^2&=&\begin{pmatrix}
m_Z^2   &-\frac{1}{3}g_X\nu m_Z(1+S_{\beta} ^2)   \\ 
&&\\
* &   m_{Z'}^2\left[1+(1+3S_{\beta}^2)\epsilon ^2\right]   \\
\end{pmatrix}.
\end{eqnarray}
By diagonalizing the above matrix, we obtain the true mass eigenstates

\begin{eqnarray}
Z_{1\mu}&=&Z_{\mu}C_{\theta}+Z'_{\mu}S_{\theta},  \nonumber \\
Z_{2\mu}&=&Z_{\mu}S_{\theta}-Z'_{\mu}C_{\theta},
\label{2-eigeneutral}
\end{eqnarray}
with the following $ZZ'$-mixing angle:

\begin{eqnarray}
S_{\theta} \approx \left(1+S_{\beta }^2\right)\epsilon _{ZZ'}^2 = (1+S_{\beta}^2)\frac{4g_X^2C_W^2}{9g^2}\left(\frac{m_Z}{m_{Z'}}\right)^2,
\end{eqnarray}
while the true mass eigenvalues are:

\begin{eqnarray}
M_{Z_1}^2&=&m_{Z}^2\left[1-\left(1+S_{\beta}^2\right)^2\epsilon _{ZZ'}^2\right], \nonumber \\ 
M_{Z_2}^2&=&m_{Z'}^2\left[1+\left(1+3S_{\beta}^2\right)\epsilon _{ZZ'}^2\right]. 
\end{eqnarray}

\subsection{The Dirac Lagrangian}

The couplings between fermions and gauge bosons, are described by the Dirac Lagrangian, which reads:

\begin{eqnarray}
\mathcal{L}_{D}=i\sum_{f,i}\overline{f^i_L}\gamma ^\mu D_{\mu} f^i_L+\overline{f^i_R}\gamma ^\mu D_{\mu}f^i_R, 
\end{eqnarray}
where $f^i_{L,R}$ contains SM and non-SM fermions. According to Tabs. \ref{tab:SM-espectro} and  \ref{tab:exotic-espectro}, the SM doublet quarks $q^i_L$ have family dependent $U(1)_X$ values. Thus, we obtain the following form:

\begin{eqnarray}
-i\mathcal{L}_{D} &=& \overline{q_L^{1}}\gamma ^{\mu}D_{\mu}q_L^1+ \overline{q_L^{a}}\gamma ^{\mu}D_{\mu}q_L^a+\overline{U_R^{i}}\gamma ^{\mu}D_{\mu}U_R^i+\overline{D_R^{i}}\gamma ^{\mu}D_{\mu}D_R^i \nonumber \\
&+&\overline{T_L}\gamma ^{\mu}D_{\mu}T_L+\overline{T_R}\gamma ^{\mu}D_{\mu}T_R +\overline{J_L^{n}}\gamma ^{\mu}D_{\mu}J_L^n +\overline{J_R^{n}}\gamma ^{\mu}D_{\mu}J_R^n  \nonumber \\
&+&\overline{\ell_L^{i}}\gamma ^{\mu}D_{\mu}\ell_L^i+\overline{e_R^{i}}\gamma ^{\mu}D_{\mu}e_R^i+\overline{(\nu _R^{i})^c}\gamma ^{\mu}D_{\mu}(\nu _R^i)^c,
\label{dirac-1}   
\end{eqnarray}
where $a=2,3$ label the second and third quark doublet and $n=1,2$ is the index of the exotic $J^n$ quarks. A sum over the indices $i,a$ and $n$ is understood. Taking into account the covariant derivative in (\ref{covariant}), the gauge mass eigenstates in (\ref{1-eigeneutral}) for the photon and (\ref{2-eigeneutral}) for the neutral weak bosons, and the quantum numbers of the fermions according to Tabs. \ref{tab:SM-espectro} and  \ref{tab:exotic-espectro}, we obtain the following terms:

\begin{eqnarray}
\mathcal{L}_{D}&=&i\sum_{f,i}\overline{f^i}\gamma ^{\mu}\partial _{\mu}f^i+\frac{g}{\sqrt{2}}\sum_i\left(\overline{U_L^i}\gamma ^{\mu}W_{\mu}^+D_L^i+\overline{\nu _L^i}\gamma ^{\mu}W_{\mu}^+e_L^i+h.c.\right) \nonumber \\
&+&e\sum_{f,i}Q_f\overline{f^i}\gamma ^{\mu}A_{\mu}f^i-\frac{g}{4C_W}\sum_{f,i}\overline{f^i}\gamma ^{\mu}\left(Z_{1\mu}C_{\theta}+Z_{2\mu}S_{\theta}\right)\left[I_f^L+I_f^R-\gamma _5(I_f^L-I_f^R)\right]f^i \nonumber \\
&-&\frac{g_X}{6}\sum_i\left[2\overline{e^i}\gamma ^{\mu}(2+\gamma _5)e^i+\overline{\nu ^i}\gamma ^{\mu}(1-\gamma _5)\nu ^i+\overline{(\nu ^i)^c}\gamma ^{\mu}(1-\gamma _5)(\nu ^i)^c\right]\left(Z_{1\mu}S_{\theta}-Z_{2\mu}C_{\theta}\right)  \nonumber \\
&-&\frac{g_X}{6}\sum_{a,n}\left[ \overline{U^1}\gamma ^{\mu}(-3-\gamma _5)U^1+2\overline{U^a}\gamma ^{\mu}(-1-\gamma _5)U^a+2\overline{D^1}\gamma ^{\mu}\gamma _5D^1+\overline{D^a}\gamma ^{\mu}(1+\gamma _5)D^a\right. \nonumber \\
&&\left. +\overline{T}\gamma ^{\mu}(-3-\gamma _5)T+\overline{J^n}\gamma ^{\mu}(1+\gamma _5)J^n\right]\left(Z_{1\mu}S_{\theta}-Z_{2\mu}C_{\theta}\right),
\label{dirac-2}
\end{eqnarray}
where $f^i$ runs over all individual fermions ($f^i=U^i,D^i,e^i,T,$ etc.),  and $e=gS_W$ is the electric charge of the proton. We show separately  the family dependent couplings in the last terms. In the above Lagrangian, we define the following parameters:

\begin{eqnarray}
 &&Q_f: 
     \begin{cases}
     Q_{U,T}=2/3, & Q_{D,T}=-1/3 \\
     Q_{e}=-1, & Q_{\nu, \nu ^c}=0
     \end{cases},
\\ \nonumber \\ \nonumber \\
&&  I_f^{L,R}=2\left(I_3^{L,R}-Q_fS_W^2\right)
 \\ \nonumber \\
  && I_3^{L,R}:
     \begin{cases}
     1/2 & \text{for the upper component of doublets: }  U_L^i, \nu _L^i  \\
     -1/2 & \text{for the lower component of doublets: }  D_L^i, e _L^i \\
     0 & \text{for any singlet component: }  f^i_R, T_L, J^n_L
     \end{cases},\label{isospin}
 \end{eqnarray}
 
In particular, for the weak neutral interaction we define the vector- and axial-couplings as two types: SM and non-SM type, where the SM type is family universal and defined as:

\begin{eqnarray}
v_f^{SM}=\frac{1}{2}\left(I_f^L+ I_f^R\right), \ \ \ a_f^{SM}=\frac{1}{2}\left(I_f^L- I_f^R\right),
\end{eqnarray}
while the non-SM couplings is family non-universal in the quark sector. Tab. \ref{tab:vector-axial-couplings} shows the individual values of the vector and axial couplings. Thus, the weak neutral Lagrangian can be written as:

\begin{eqnarray}
\mathcal{L}_{WN}&=&Z_{1\mu }\sum _{f,i}\overline{f^i}\gamma ^{\mu}\left[\frac{g}{2C_W}(v_f^{SM}-\gamma _5a_f^{SM})C_{\theta}+\frac{g_X}{2}(v_i^{NSM}-\gamma _5a_i^{NSM})S_{\theta}\right]f^i \nonumber \\
&+&Z_{2\mu }\sum _{f,i}\overline{f^i}\gamma ^{\mu}\left[\frac{g}{2C_W}(v_f^{SM}-\gamma _5a_f^{SM})S_{\theta}-\frac{g_X}{2}(v_i^{NSM}-\gamma _5a_i^{NSM})C_{\theta}\right]f^i, 
\label{weak current}
\end{eqnarray}
where a small non-SM coupling with the lighter weak boson $Z_1$ arises. This {\it exotic} coupling disappears if we ignore the mixing angle (i.e. $S_{\theta}=0$), obtaining the usual SM weak neutral currents.   

\subsection{Yukawa Lagrangian}

We find the Yukawa Lagrangian compatible with the $G_{sm}\times U(1)_X$ symmetry. For the quark sector we find:

\begin{eqnarray}
-\mathcal{L}_Q &=& \overline{q_L^{1}}\left(\widetilde{\phi} _2h^{U}_2 \right)_{1j}U_R^{j}+\overline{q_L^{a}}(\widetilde{\phi }_1 h^{U}_{1})_{aj}U_R^{j}+\overline{q_L^{1}}\left(\phi _1 h^{D}_1\right)_{1j}D_R^{j}+\overline{q_L^{a}}\left(\phi _2 h^{D}_{2} \right)_{aj}D_R^{j} \notag \\
&+&\overline{q_L^{1}} (\phi _1 h^{J}_{1})_{1m} J^{m}_R+\overline{q_L^{a}}\left(\phi  _2 h^{J}_{2} \right)_{am} J^{m}_R+\overline{q_L^{1}}\left(\widetilde{\phi} _2 h^{T}_{2} \right)_1T_R +\overline{q_L^{a}} (\widetilde{\phi } _1 h^{T}_{1})_aT_R \notag \\
&+&\overline{T_{L}}\left( \sigma_0^* h_{\sigma }^{U}+\chi _0^*h_{\chi }^{U}\right)_{j}{U}_{R}^{j}+\overline{T_{L}}\left( \sigma_0^*h_{\sigma}^{T}+\chi _0^*h_{\chi }^{T}\right){T}_{R}
\nonumber \\
&+&\overline{J_{L}^n}\left( \sigma_0h_{\sigma }^{D}+\chi _0h_{\chi }^{D}\right)_{nj}{D}_{R}^{j}+\overline{J_{L}^n}\left( \sigma_0h_{\sigma }^{J}+\chi _0h_{\chi }^{J}\right)_{nm}{J}_{R}^{m}+h.c.,
 \label{quark-yukawa-1}
\end{eqnarray}
where $\widetilde{\phi}_{1,2}=i\sigma_2 \phi_{1,2}^*$ are conjugate fields, 
For the leptonic sector we obtain:

\begin{eqnarray}
-\mathcal{L}_{\ell}&=&\overline{\ell ^{i}_{L}}\left( \widetilde{\phi}_1h_{1 }^{\nu } \right)_{ij}{\nu }_{R}^{j}+ \overline{\ell ^{i}_{L}}\left( \widetilde{\phi}_2h_{2 }^{N}\right)_{ij}{N}_{R}^{j}\nonumber \\
&+&\overline{(\nu ^{i}_{R})^c}\left( \sigma _0^*h_{\sigma }^{N}+\chi _0^* h_{\chi }^{N}\right)_{ij}{N}_{R}^j +\frac{1}{2}M_{N}\overline{(N^{i}_{R})^c}{N}_{R}^{j}\nonumber \\
&+&\overline{\ell ^{i}_{L}}\left( \phi _1h_{1}^{e } \right)_{ij}{e}_{R}^{j}+h.c.
\label{yukawa-leptons-1}
\end{eqnarray}

In particular, we can see in the quark Lagrangian in Eq. (\ref{quark-yukawa-1}) that due to the non-universality of the $U(1)_X$ symmetry, 
not all couplings between quarks and scalars are allowed by the gauge symmetry, which lead us to specific zero-texture Yukawa matrices. To see this, we write  (\ref{quark-yukawa-1}) in a shorter form as:

\begin{eqnarray}
-\mathcal{L}_Q&=&\overline{q^{i}_{L}}\left( \widetilde{\phi}_1h_{1}^{U}+\widetilde{\phi}_2 h_{2}^{U}\right)_{ij}{U}_{R}^{j}+\overline{q^{i}_{L}}\left( \phi_1h_{1}^{D}+\phi _2h_{2}^{D}\right)_{ij}{D}_{R}^{j} \nonumber \\
&+&\overline{q^{i}_{L}}\left( \phi_1h_{1}^{J}+\phi _2h_{2}^{J}\right)_{im}{J}_{R}^{m}+\overline{q^{i}_{L}}\left( \widetilde{\phi}_1h_{1}^{T}+\widetilde{\phi}_2 h_{2 }^{T}\right)_i{T}_{R} \nonumber \\
&+&\overline{T_{L}}\left( \sigma_0^*h_{\sigma }^{U}+\chi _0^*h_{\chi }^{U}\right)_{j}{U}_{R}^{j}+\overline{T_{L}}\left( \sigma_0^*h_{\sigma}^{T}+\chi _0^*h_{\chi }^{T}\right){T}_{R}
\nonumber \\
&+&\overline{J_{L}^n}\left( \sigma_0h_{\sigma }^{D}+\chi _0h_{\chi }^{D}\right)_{nj}{D}_{R}^{j}+\overline{J_{L}^n}\left( \sigma_0h_{\sigma }^{J}+\chi _0h_{\chi }^{J}\right)_{nm}{J}_{R}^{m}+h.c.,
\label{yukawa-quarks-2}
\end{eqnarray}   
where the Yukawa matrices exhibit the following zero-textures:

\begin{eqnarray}
h_{1}^{U}&=&\begin{pmatrix}
0 & 0 & 0 \\
a_{21} & a_{22} & a_{23} \\
a_{31} & a_{32} & a_{33} \\
\end{pmatrix}, \hspace{0.3cm}
h_{2}^{U}=\begin{pmatrix}
b_{11} & b_{12} & b_{13} \\
0 & 0 & 0 \\
0 & 0 & 0 \\
\end{pmatrix},\nonumber
\\ \nonumber \\
h_{1}^{D}&=&\begin{pmatrix}
A_{11} & A_{12} & A_{13} \\
0 & 0 & 0 \\
0 & 0 & 0 \\
\end{pmatrix}, \hspace{0.3cm}
h_{2}^{D}=\begin{pmatrix}
0 & 0 & 0 \\
B_{21} & B_{22} & B_{23} \\
B_{31} & B_{32} & B_{33} \\
\end{pmatrix},  
 \notag
\end{eqnarray}
\begin{eqnarray}
h_{\sigma, \chi}^{D}&=&
\begin{pmatrix}
C_{11} & C_{12} & C_{13} \\
C_{21} &  C_{22} & C_{23} \\
\end{pmatrix}, \hspace{0.3cm} h_{\sigma, \chi}^{U}=
(c_1, c_2,c_3),  \nonumber
\\ \nonumber \\
h_{1}^{J}&=&\begin{pmatrix}
j_{11} & j_{12}\\
0  & 0\\
0 & 0 \\
\end{pmatrix}, \hspace{0.3cm}
h_{2}^{J}=\begin{pmatrix}
0 & 0\\
i_{21} & i_{22}\\
i_{31} & i_{32} \\
\end{pmatrix},  \hspace{0.3cm}
h_{\sigma, \chi}^{J}=
\begin{pmatrix}
k _{11} & k_{12}\\
k_{21} & k_{22} \\
\end{pmatrix}, \nonumber \\ \nonumber \\ 
h_{1}^{T}&=&\begin{pmatrix}
0 \\
w_2  \\
w_3  \\
\end{pmatrix}, \hspace{0.3cm}
h_{2}^{T}=\begin{pmatrix}
y_1 \\
0 \\
0 \\
\end{pmatrix},  \hspace{0.3cm}
h_{\sigma, \chi}^{T}=h_T.
\label{quark-textures}
\end{eqnarray}

\section{Phenomenology}

To explore the consequences of the above couplings, we consider some phenomenological predictions.
 
\subsection{Z' production}

\begin{figure}[tbh]
\centering
\includegraphics[width=12cm,height=10cm,angle=0]{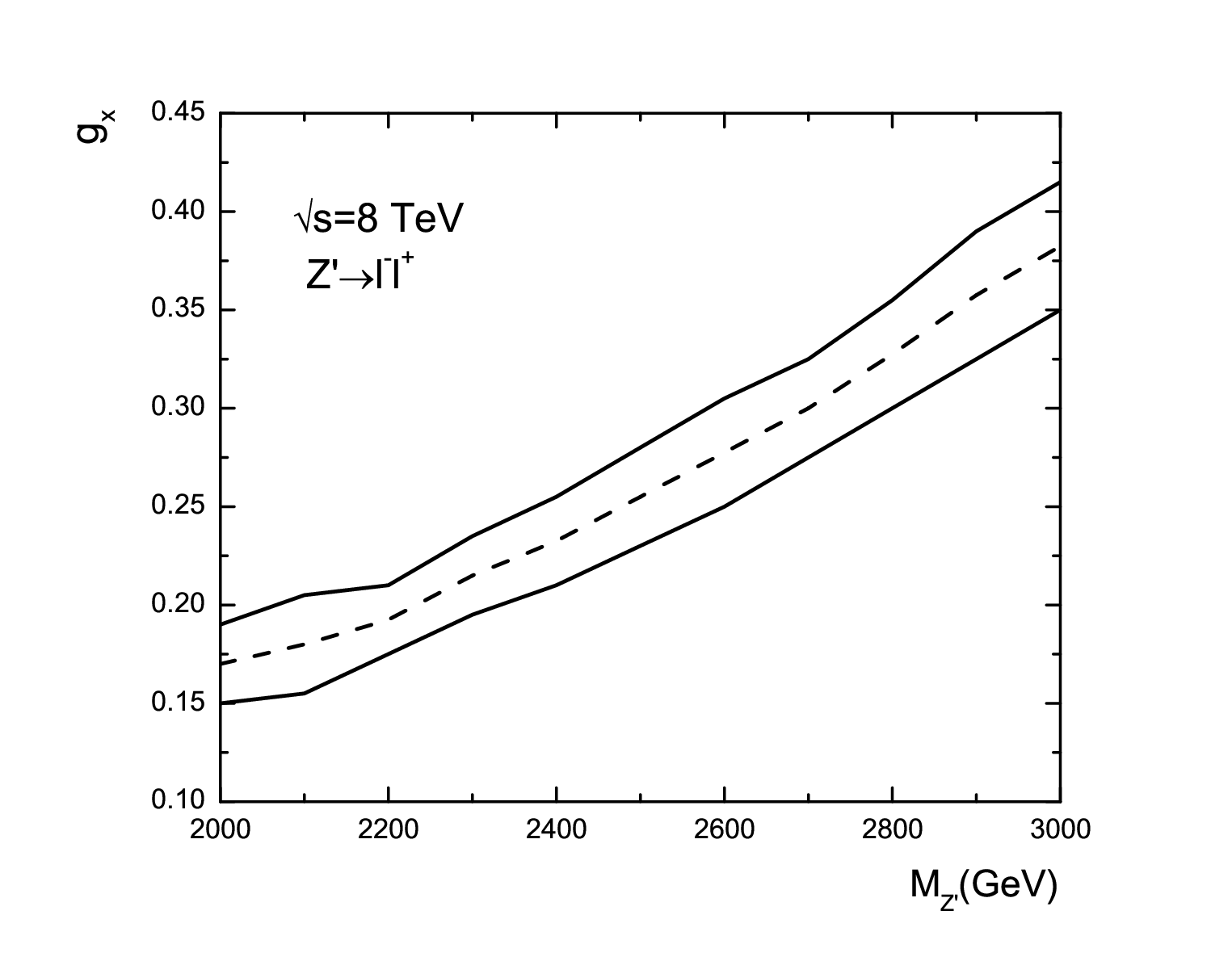}
\caption{Superior limit and its $2\sigma$ band in the $g_X-M_{Z'}$ plane obtained by comparing the simulated $pp\rightarrow Z' \rightarrow \ell ^+ \ell ^-$ cross section and the $95\%$ C.L. bounds found at LHC in the searching for $Z'$ resonances.}
\label{fig1}
\end{figure}

We consider the physics of the extra neutral gauge boson at high energy, where the mixing $Z-Z'$ angle is not very sensitive. Thus, we assume that $S_{\theta}=0$ in the coupling with $Z_{2\mu }$ in (\ref{weak current}), from where $Z_2=Z'$. The principal $Z'$ production processes in proton-proton collisions are Drell-Yan processes with two-body final states. In the resonance, the cross section depends on two free parameters: the $U(1)_X$ coupling constant $g_X$ and the $Z'$ mass. In order to obtain some constraints on these parameters, we use the experimental limits on Drell-Yan cross section recently obtained at the CERN-LHC collider by the ATLAS collaboration \cite{cross-sect}  at $8$ TeV center-of-mass energy, where cross section limits vs. $m_{Z'}$ were obtained at 95\% CL in dilepton resonance, obtaining no significant deviation from the SM expectation into the $\pm 2\sigma$ bound. Using the CalcHep package, we simulate $pp \rightarrow Z' \rightarrow e^+e^- (\mu ^+\mu ^-)$ events in the framework of our model and compare with the experimental bounds presented in \cite{cross-sect} in  the range between $m_{Z'}=2$ and $3$ TeV. Fig. \ref{fig1} shows upper limits in the $g_X-m_{Z'}$ plane and its $2\sigma$ band, obtaining maximum values as low as $g_X=0.15$ for $m_Z'=2$ TeV, to values as large as $g_X=0.42$ at  $m_Z'=3$ TeV.



On the other hand, due to the large production rates of top-quarks at the LHC collider, it is possible to explore the high $t\overline{t}$ invariant mass region to search for new physics as new resonances, or  new couplings that causes anomalies at lower energies, as for instance, the large forward-backward asymmetry (FBA) in top quark pairs observed at Tevatron \cite{FBA}. We consider the invariant-mass distribution in the process $pp \rightarrow Z' \rightarrow t\overline{t}$. For top-quarks, we use the following kinematical and dynamical parameters:

\begin{itemize}
\item[-] seudorapidity within $|\eta| < 2.5 $
\item[-] Transverse momentum $p_T > 200 GeV$
\item[-] Proton-proton collisions at center-of-mass energy of 8 TeV.
\item[-] top-quark mass of 172.5 GeV
\end{itemize}

\begin{figure}[tbh]
\centering
\includegraphics[width=12cm,height=10cm,angle=0]{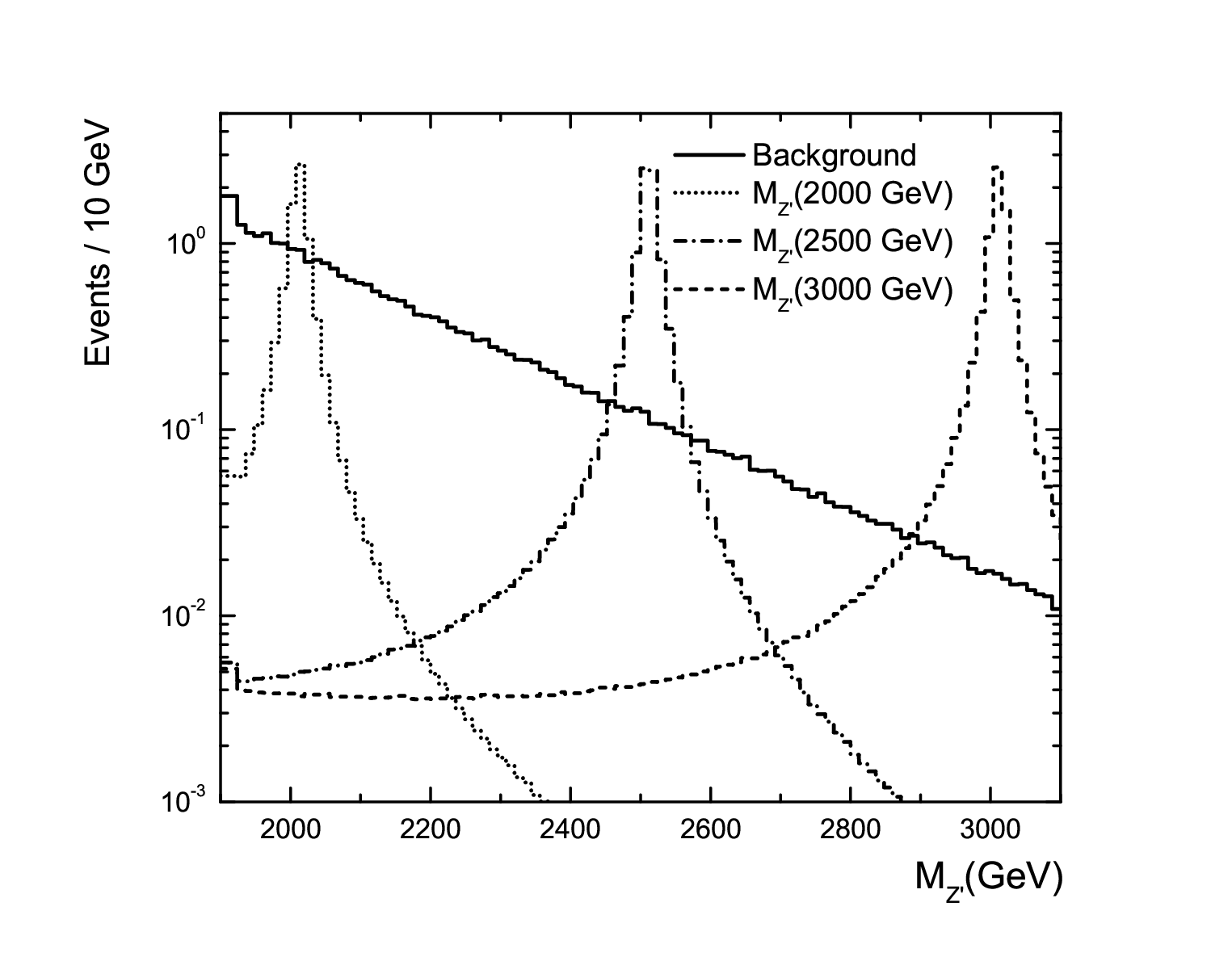}
\caption{Invariant-mass distribution for three $Z'$ mass values: 2, 2.5 and 3 TeV, for $g_X=0.19, 0.3$ and $0.42$, respectively.   }
\label{fig2}
\end{figure}

\begin{figure}[tbh]
\centering
\includegraphics[width=12cm,height=10cm,angle=0]{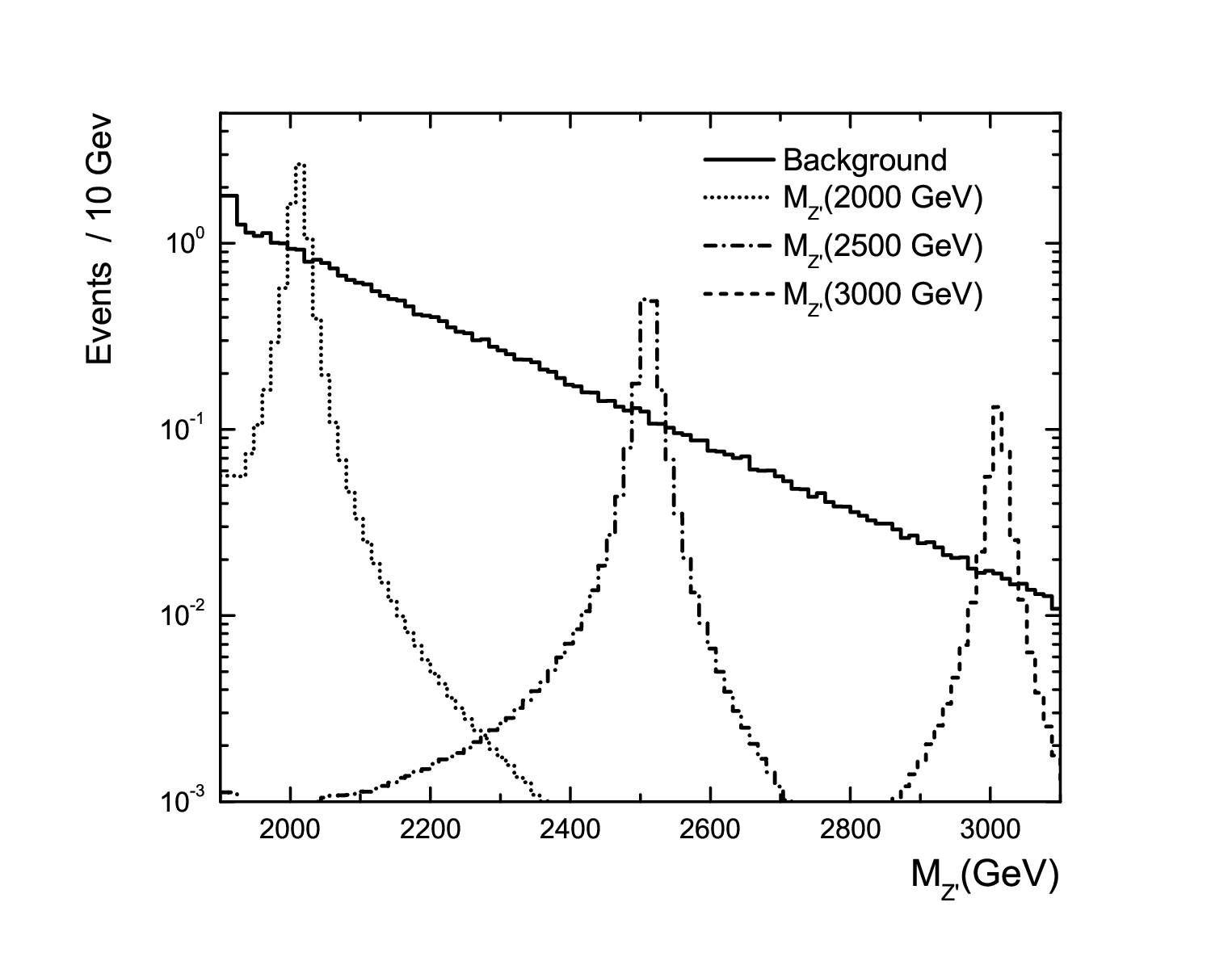}
\caption{Invariant-mass distribution for three $Z'$ mass values: 2, 2.5 and 3 TeV, for $g_X=0.19$.}
\label{fig3}
\end{figure}

Taking into account the normal assignation in (\ref{quark-assignation}), the couplings from (\ref{weak current}) lead to the following partial width of $Z'$ into top pairs

\begin{eqnarray}
\Gamma_{Z' \rightarrow t\overline{t}}=\frac{g_X^2m_{Z'}}{16\pi }\sqrt{1-\mu _{t}'^{2}}\left[\left(1+\mu _t'^{2}/2\right)\left(v_{t}^{NSM}\right)^2+\left(1-\mu _t'^{2}\right)\left(a_t^{NSM}\right)^2\right],
\label{zprime-width}
\end{eqnarray}
with $\mu _t'^{2}=4m_{t}^2/m_{Z'}^2$. First, by choosing variable limits for the $g_X$ and $Z'$ mass values according to Fig. \ref{fig1}, we obtain the invariant-mass distribution based on an integrated luminosity $L=100$ $fb^{-1}$, which is shown in Fig. \ref{fig2} for $m_{Z'}= 2, 2.5$ and $3$ TeV, with $g_X=0.19, 0.3$ and $0.42$, respectively. Since the limits of the coupling constant $g_X$ increase with the $Z'$ mass, the resonances exhibit about the same values for the three proved $Z'$ masses, while the SM background decreases with the invariant-mass. On the other hand, if we choose one fixed value for the $g_X$ coupling, we obtain different mass distribution for each $Z'$ mass. For example, Fig. \ref{fig3} shows the three $Z'$ resonances for $g_X=0.19$, which decrease with the mass, but still with a small excess over the background.    




\subsection{Quark masses} 
Due to the zero-texture structure exhibit by the Yukawa couplings in (\ref{yukawa-quarks-2}), and taking into account normal assignation according to (\ref{quark-assignation}), the model can reproduce the mass hierarchy in the quark sector. After the symmetry breaking, we obtain the following mass terms:

\begin{eqnarray}
-\langle\mathcal{L}_Q \rangle&=&\overline{U^{i}_{L}}\left( \nu _1h_{1}^{U}+\nu _2h_{2}^{U}\right)_{ij}{U}_{R}^{j}+\overline{D^{i}_{L}}\left(\nu _1h_{1}^{D}+\nu _2h_{2}^{D}\right)_{ij}{D}_{R}^{j} \nonumber \\
&+&\overline{U^{i}_{L}}\left( \nu _1h_{1}^{T}+\nu _2 h_{2 }^{T}\right)_i{T}_{R}+\overline{D^{i}_{L}}\left( \nu _1h_{1}^{J}+\nu _2h_{2}^{J}\right)_{im}J^m_R \nonumber \\
&+&\overline{T_{L}}\left( \nu _{\sigma }h_{\sigma }^{U}+\nu _{\chi}h_{\chi }^{U}\right)_{j}U_{R}^{j}+\overline{T_{L}}\left( \nu _{\sigma }h_{\sigma }^{T}+\nu _{\chi }h_{\chi }^{T}\right){T}_{R}
\nonumber \\
&+&\overline{J_{L}^n}\left( \nu _{\sigma }h_{\sigma }^{D}+\nu _{\chi }h_{\chi }^{D}\right)_{nj}{D}_{R}^{j}+\overline{J_{L}^n}\left( \nu _{\sigma }h_{\sigma }^{J}+\nu _{\chi } h_{\chi }^{J}\right)_{nm}{J}_{R}^{m}+h.c.
\label{mass-quarks-1}
\end{eqnarray} 

On the other hand, since (\ref{yukawa-quarks-2}) exhibits terms where $\phi _1$ and $\phi _2$ or $\sigma _0$ and $\chi _0$ couple simultaneously, the zero structures of the Yukawa matrices in (\ref{quark-textures}) does not imply zero-texture mass matrices. Thus, the extra $U(1)_X$ symmetry is not sufficient to explain the mass spectrum. 
Thus, we assume the existence of two types of global symmetries. They are:

\begin{itemize}
\item {\it $Z_2$ symmetries:} We restrict the couplings of scalar fields by requiring the discrete symmetries

\begin{eqnarray}
\phi _2 &\rightarrow& -\phi _2, \hspace{1cm} \sigma _0 \rightarrow -\sigma _0, \hspace{1cm}D_R^i \rightarrow -D_R^i, \hspace{1cm} T_{L,R} \rightarrow -T_{L,R}.
\label{discrete}
\end{eqnarray} 



\item  {\it $U(1)_{T_3}$ symmetry:} The $U(1)_X$ symmetry distinguishes the quark family $q^1_L$ from the others two $q^{a=2,3}_L$, while the right-handed components are universal. Thus, in the absence of the Yukawa couplings, the model has the following global symmetry:

\begin{eqnarray}
G_{global}(h^Q=0)=SU(2)_{q^a}\times SU(3)_{U^i} \times SU(3)_{D^i}.
\label{global}
\end{eqnarray}
In particular, the $SU(2)_{q^a}$ symmetry in the left-handed sector remains in the model even after the gauge symmetry breaking. However, the experimental observation shows that this symmetry does not remain if the quark masses are taken into account. Let us assume that the Yukawa interactions break the $SU(2)_{q^a}$ global symmetry, but an $U(1)_{T_3}$ symmetry remains only in the left-handed down sector, under which





\begin{eqnarray}
D_L^2\rightarrow -D_L^2,  \hspace{1cm} D_L^3 \rightarrow D_L^3.
\label{discrete-2}
\end{eqnarray}
    
\end{itemize} 

Thus, by requiring the symmetries (\ref{discrete}) and (\ref{discrete-2}), the mass Lagrangian (\ref{mass-quarks-1}) becomes:

\begin{eqnarray}
-\langle\mathcal{L}_Q \rangle&=&\overline{U^{i}_{L}}\left( \nu _1h_{1}^{U}\right)_{ij}{U}_{R}^{j}+\left[\overline{D^{1}_{L}}\left(\nu _2h_{2}^{D}\right)_{1j}+\overline{D^{2}_{L}}\left(\nu _1h_{1}^{D}\right)_{2j}+\overline{D^{3}_{L}}\left(\nu _2h_{2}^{D}\right)_{3j}\right]{D}_{R}^{j} \nonumber \\
&+&\overline{U^{i}_{L}}\left(\nu _2 h_{2 }^{T}\right)_i{T}_{R}+\left[\overline{D^{1}_{L}}\left( \nu _1h_{1}^{J}\right)_{1m}+\overline{D^{2}_{L}}\left( \nu _2h_{2}^{J}\right)_{2m}+\overline{D^{3}_{L}}\left( \nu _1h_{1}^{J}\right)_{3m} \right]J^m_R \nonumber \\
&+&\overline{T_{L}}\left( \nu _{\sigma }h_{\sigma }^{U}\right)_{j}U_{R}^{j}+\overline{T_{L}}\left( \nu _{\chi }h_{\chi }^{T}\right){T}_{R}+\overline{J_{L}^n}\left( \nu _{\sigma }h_{\sigma }^{D}\right)_{nj}{D}_{R}^{j}+\overline{J_{L}^n}\left( \nu _{\chi } h_{\chi }^{J}\right)_{nm}{J}_{R}^{m}+h.c. \nonumber \\ \nonumber \\
-\langle\mathcal{L}_Q \rangle&=&\overline{U^{i}_{L}}(M_{U})_{ij}{U}_{R}^{j}+\overline{D^{i}_{L}}(M_{D})_{ij}{D}_{R}^{j}+\overline{U^{i}_{L}}(k)_i{T}_{R} +\overline{D^{i}_{L}}(s)_{im}J^m_R
\nonumber \\
&+&\overline{T_{L}}(K)_{j}U_{R}^{j}+\overline{T_{L}}(M_T){T}_{R}+\overline{J_{L}^n}(S)_{nj}{D}_{R}^{j}+\overline{J_{L}^n}(M_{J})_{nm}{J}_{R}^{m}+h.c.
\label{mass-quarks-3}
\end{eqnarray}
where the mass matrices are:

\begin{eqnarray}
M_{U}&=&\frac{\nu _1}{\sqrt{2}}\begin{pmatrix}
0 & 0 & 0 \\
a_{21} & a_{22} & a_{23} \\
a_{31} & a_{32} & a_{33} \\
\end{pmatrix}, \hspace{0.5cm}
M_{D}=\frac{\nu _2}{\sqrt{2}}\begin{pmatrix}
0 & 0 & 0 \\
0 & 0 & 0 \\
B_{31} & B_{32} & B_{33} \\
\end{pmatrix},  \nonumber \\ \nonumber \\
M_J&=&\frac{\nu _{\chi }}{\sqrt{2}}\begin{pmatrix}
k _{11} & k_{12}\\
k_{21} & k_{22} \\
\end{pmatrix}, \hspace{0.5cm}
M_T=\frac{\nu _{\chi}}{\sqrt{2}}h_{\chi }^T, \nonumber \\ \nonumber \\
S&=&\frac{\nu _{\sigma }}{\sqrt{2}}\begin{pmatrix}
C_{11} & C_{12} & C_{13} \\
C_{21} &  C_{22} & C_{23} \\
\end{pmatrix}, \hspace{0.5cm} K=\frac{\nu _{\sigma}}{\sqrt{2}}(c_1, c_2,c_3),\nonumber \\ \nonumber \\
s&=&
\frac{\nu _1}{\sqrt{2}}\begin{pmatrix}
j_{11} & j_{12}\\
0  & 0\\
0 & 0 \\
\end{pmatrix}+\frac{\nu _2}{\sqrt{2}}\begin{pmatrix}
0 & 0\\
i_{21}  & i_{22}\\
0 & 0 \\ 
\end{pmatrix}, \hspace{0.8cm} k=\frac{\nu _2}{\sqrt{2}}\begin{pmatrix}
y_1 \\
0 \\
0 \\
\end{pmatrix}.
\label{mass-matrices-1}
\end{eqnarray}

We see that in absence of mixing between the ordinary and the exotic sector, the matrices of the SM quarks have the form

\begin{eqnarray}
M_{U}=\frac{\nu _1}{\sqrt{2}}h^{U}= \begin{pmatrix}
0 && 0 && 0 \\
* && * && * \\
* && * && * \\
\end{pmatrix},
\hspace{1cm}
 M_{D}=\frac{\nu _2}{\sqrt{2}}h^{D}= \begin{pmatrix}
0 && 0 && 0 \\
0 && 0 && 0 \\
* && * && * \\
\end{pmatrix}.
\label{nonsm-matrix}
\end{eqnarray} 
which exhibit three massless ($m_{u,d,s}^0=0$) and three massive ($m_{c,b,t}^0 \sim \nu _{1,2} \sim$ GeV) quarks, while the non-SM quarks acquire heavy mass values ($m_{T,J}^0 \sim \nu _{\chi} \gg \nu$). However, if we consider the contributions due to the mixing mass matrices $s, k, S$ and $K$, the extended mass matrices are  

\begin{eqnarray}
M'_{U}&=& \left( 
\begin{array}{ccc}
M_U  &\left| \right. &k \\
\text{\textemdash \hspace{0.1cm} \textemdash} & \text{\textemdash}  &  \text{\textemdash  \hspace{0.1cm} \textemdash} \\ 
K  &\left| \right. & M_T%
\end{array}%
\right)=\frac{1}{\sqrt{2}}\begin{pmatrix}
0 & 0 & 0 &  \left| \right. & \nu _2y_1   \\
\nu _1a_{21} & \nu _1a_{22} & \nu _1a_{23} &  \left| \right. & 0 \\
\nu _1a_{31} & \nu _1a_{32} & \nu _1a_{33} & \left| \right. & 0\\
\text{\textemdash} & \text{\textemdash} & \text{\textemdash} & \text{\textemdash} & \text{\textemdash}\\
\nu _{\sigma}c_1  & \nu _{\sigma}c_2  & \nu _{\sigma}c_3   &  \left| \right.  & \nu _{\chi }h_{\chi }^T
\end{pmatrix}, \nonumber
\\  \nonumber 
\\
 M'_{D}&=&\left( 
\begin{array}{ccc}
M_D & \left| \right. &s \\
\text{\textemdash \hspace{0.1cm} \textemdash} & \text{\textemdash}  &  \text{\textemdash  \hspace{0.1cm} \textemdash} \\ 
S & \left| \right. & M_J%
\end{array}%
\right)=\frac{1}{\sqrt{2}}\begin{pmatrix}
0 & 0 & 0 &  \left| \right.  & \nu _1j_{11} & \nu _1j_{12}   \\
0 & 0 & 0 &  \left| \right.  & \nu _2i_{21} & \nu _2i_{22} \\
\nu _2B_{31} & \nu _2B_{32} & \nu _2B_{33} & \left| \right.  & 0 &0 \\
 \text{\textemdash}  &  \text{\textemdash}  &  \text{\textemdash}  &  \text{\textemdash}  &  \text{\textemdash} &  \text{\textemdash}   \\
\nu _{\sigma }C_{11}  & \nu _{\sigma }C_{12}   & \nu _{\sigma }C_{13}   &  \left| \right.  & \nu _{\chi }k_{11} & \nu _{\chi }k_{12} \\
\nu _{\sigma }C_{21}  & \nu _{\sigma }C_{22}  & \nu _{\sigma }C_{23}  &  \left| \right. & \nu _{\chi }k_{21}  & \nu _{\chi }k_{22}
\end{pmatrix}.
\label{mass-matrices-2}
\end{eqnarray} 
which exhibit non-vanishing determinant, providing masses to all quarks.  
Thus, due to the mixing components, the mass matrices $M_{U,D}'$ exhibits three eigenvalues at the scale $m_{u,d,s} \sim$ MeV, three at the scale $m_{c,b,t} \sim$ GeV and three at the scale $m_{T,J} \sim$ TeV.

To explore the consequences of the above mass scheme, we consider an specific simple structure of the matrices in (\ref{mass-matrices-2}) from the naturalness criterion that the Yukawa couplings have similar values to each other. To achieve this without spoil the mass structures, we assume the scenery where the mixing terms are diagonal ($c_{2,3}=k_{ij}=j_{ij}=C_{ij}=0$ for $i \neq j$, while  $\nu _1j_{11}=\nu _2i_{22}=h_D$ and $C_{11}=C_{22}=\Gamma _D $), the $M_U$ sector have identical Yukawa components except the top coupling (i.e. $a_{ij}=Y_U$ for $ij \neq 33$ and $a_{33} = Y_t$); and $M_D$ have identical components (i.e. $B_{31}=B_{32}=B_{33}=Y_D$). Thus, the matrices in (\ref{mass-matrices-2}) become:


\begin{eqnarray}
M'_{U}&=& \frac{1}{\sqrt{2}}\begin{pmatrix}
0 & 0 & 0 &  \left| \right. & \nu _2y_1   \\
\nu _1Y_U & \nu _1Y_U & \nu _1Y_U &  \left| \right. & 0 \\
\nu _1Y_U & \nu _1Y_U & \nu _1Y_t & \left| \right. & 0\\
\text{\textemdash} & \text{\textemdash} & \text{\textemdash} & \text{\textemdash} & \text{\textemdash}\\
\nu _{\sigma}c_1  & 0  & 0   &  \left| \right.  & \nu _{\chi }h_{\chi }^T
\end{pmatrix}, \nonumber
\\  \nonumber 
\\
 M'_{D}&=&\frac{1}{\sqrt{2}}\begin{pmatrix}
0 & 0 & 0 &  \left| \right.  & h_{D} & 0   \\
0 & 0 & 0 &  \left| \right.  & 0 & h_{D} \\
\nu _2Y_{D} & \nu _2Y_{D} & \nu _2Y_{D} & \left| \right.  & 0 &0 \\
 \text{\textemdash}  &  \text{\textemdash}  &  \text{\textemdash}  &  \text{\textemdash}  &  \text{\textemdash} &  \text{\textemdash}   \\
\nu _{\sigma }\Gamma _{D}  & 0   & 0   &  \left| \right.  & \nu _{\chi }k_{11} & 0 \\
0  & \nu _{\sigma }\Gamma _{D}   & 0  &  \left| \right. & 0  & \nu _{\chi }k_{22}
\end{pmatrix}.
\label{mass-matrices-3}
\end{eqnarray} 

The above matrices are diagonalized through bi-unitary transformation of the form $m_Q=(\mathcal{O}^Q_L)^{\dagger}M'_Q\mathcal{O}^Q_R$, with $m_Q$ a diagonal matrix with real and positive values. In order to guarantee real values, we calculate the squared matrices $M_Q^{'2}=M'_QM_Q^{'\dagger}$ which diagonalize as $m_Q^2=(\mathcal{O}^Q_L)^{\dagger}M_Q^{'2}\mathcal{O}^Q_L$. We find for the up sector the following approximate eigenvalues:

\begin{eqnarray}
\lambda _1^U&=&m_u
\approx \left(\frac{y_1c_1\nu _{\sigma}\nu _2}{\sqrt{2}h_{\chi }^T\nu _{\chi }}\right)=y_{1}c_1\left(\frac{\nu _2\nu _{\sigma}}{2m _{T}}\right) \nonumber \\
\lambda _2^U&=&m_c\approx 
\frac{\nu _1}{2\sqrt{2}}(Y_U+Y_t)\left[1-\sqrt{1+4y_{Ut}\epsilon _{Ut}}\right] \ \nonumber \\
\lambda _3^U&=&m_t\approx
\frac{\nu _1}{2\sqrt{2}}(Y_U+Y_t)\left[1+\sqrt{1+4y_{Ut}\epsilon _{Ut}}\right] \nonumber \\
\lambda _4^U&=&m_{T}\approx \frac{1}{\sqrt{2}}h_{\chi }^T\nu _{\chi },
\label{up-masses}
\end{eqnarray} 
where we consider that $\nu _{\chi} \gg \nu _{2}, \nu _{\sigma }$ and define the parameters

\begin{equation}
y_{Ut}=\frac{Y_U}{Y_U+Y_t}, \hspace{1cm} \epsilon _{Ut}=\frac{Y_U-Y_t}{Y_U+Y_t}.
\label{epsilon}
\end{equation}
The parameter $\epsilon _{Ut}$ "measures" the level of asymmetry of Yukawa interactions between the top quark and the lighter ones (charm and up). It is interesting to note that if $\epsilon _{Ut} =0$ (i.e. $Y_U=Y_t$), we obtain interactions with an exact flavor symmetry between flavors $a=2,3$. As a consequence, the c-quark becomes massless.  Furthermore, the ratio between the mass of the c- and t-quark is sensible to the asymmetry parameter according to: 

\begin{equation}
\frac{m_c}{m_t}\approx \frac{-y_{Ut}\epsilon _{Ut}}{1+y_{Ut}\epsilon _{Ut}}.
\label{ct-ratio}
\end{equation}
Thus, there is a scenery where the closer are the Yukawa values $Y_U$ and $Y_t$, the larger is the difference between $m_c$ and $m_t$. 

For the down sector we find: 

\begin{eqnarray}
\lambda _1^D&=&m_d\approx 
\left(\frac{\Gamma _Dh_D\nu _{\sigma}}{\sqrt{2}k_{11}\nu _{\chi }}\right)=j_{12}\Gamma _D\left(\frac{\nu _1\nu _{\sigma}}{2m _{J^1 }}\right) \nonumber \\
\lambda _2^D&=&m_s\approx 
\left(\frac{\Gamma _Dh_D\nu _{\sigma}}{\sqrt{2}k_{22}\nu _{\chi }}\right)=j_{12}\Gamma _D\left(\frac{\nu _1\nu _{\sigma}}{2m _{J^2 }}\right)   \nonumber \\
\lambda _3^D&=&m_b\approx \frac{1}{\sqrt{2}}Y_D\nu _2 \nonumber \\
\lambda _4^D&=&m_{J^1}\approx \frac{1}{\sqrt{2}}k_{11}\nu _{\chi }, \nonumber \\
\lambda _5^D&=&m_{J^2}\approx \frac{1}{\sqrt{2}}k_{22}\nu _{\chi }.
\label{down-masses}
\end{eqnarray}  
In this case, the ratio between the masses of the down and strange quarks gives:

\begin{equation}
\frac{m_d}{m_s}=\frac{m_{J^2}}{m_{J^1}}.
\label{ds-ratio}
\end{equation}
Thus, the ratio between the lightest quarks is determined only by the mass splitting of the heavy quarks $J^1$ and $J^2$. Regarding $m_u$ and $m_b$, we find that:

\begin{equation}
\frac{m_u}{m_b}=\left(\frac{y_1c_1}{\sqrt{2}Y_D}\right)\frac{\nu _{\sigma }}{m_T}.
\label{ub-ratio}
\end{equation}

Considering the central values, the experimental masses of the phenomenological quarks are \cite{data}:

\begin{eqnarray}
m_u&=&2.3 \text{ MeV},\hspace{1cm} m_d=4.8 \text{ MeV}, \hspace{1cm}  m_s=95 \text{ MeV}, \nonumber \\
m_c&=&1.275 \text{ GeV}, \hspace{0.7cm} m_b=4.65 \text{ GeV}, \hspace{1cm}  m_t=173.5 \text{ GeV}
\label{experimental-mass} 
\end{eqnarray}
Using the above values, the relations (\ref{ct-ratio}), (\ref{ds-ratio}) and  (\ref{ub-ratio})  leads to:

\begin{eqnarray}
y_{Ut}\epsilon _{Ut}&=&\dfrac{-m_c/m_t}{1+m_c/m_t}\approx -7,3 \times 10^{-3} \label{numeric1} \\
m_{J^1} &\approx & 20\text{ }m_{J^2} \label{numeric2} \\
\frac{\nu _{\sigma }}{m_T} &\approx &\left(5 \times 10^{-4}\right)\frac{\sqrt{2}Y_D}{y_1c_1}.
\label{numeric3}
\end{eqnarray} 

\begin{figure}[tbh]
\centering
\includegraphics[width=12cm,height=7cm,angle=0]{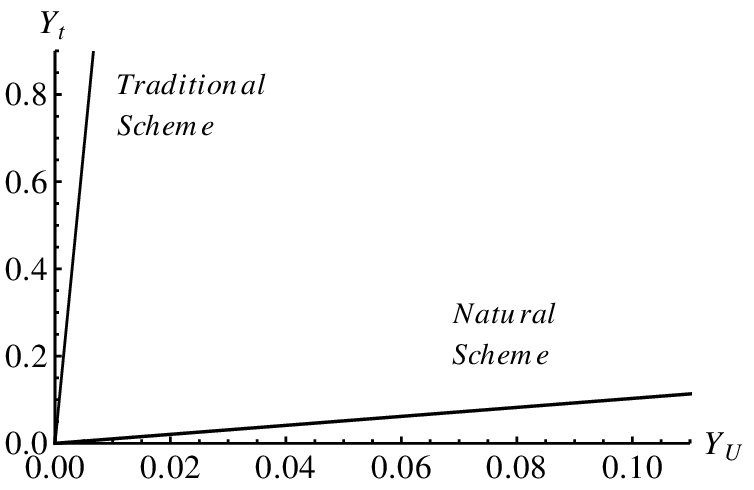}
\caption{Yukawa coupling relation of top ($Y_t$) and light ($Y_U$) quarks compatible with the experimental ratio $m_c/m_t=7.35 \times 10^{-3}$. Both lines show two schemes: the {\it traditional scheme} with $Y_t/Y_U \approx 135.05$ and the {\it natural scheme} with $Y_t/Y_U \approx 1.03$.}
\label{fig4}
\end{figure}

\begin{figure}[tbh]
\centering
\includegraphics[width=12cm,height=7cm,angle=0]{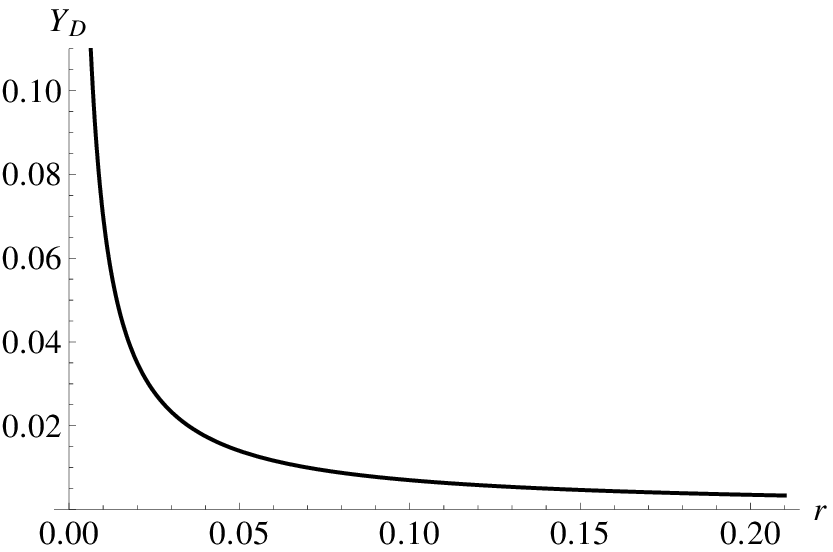}
\caption{Yukawa coupling of down-type quarks $Y_D$ as function of the ratio $r=\nu _{\sigma}/m_T$ compatible with the experimental ratio $m_u/m_b=5 \times 10^{-4}$.}
\label{fig5}
\end{figure}

Fig. \ref{fig4} shows the top quark coupling $Y_t$ as function of the light-quark coupling $Y_U$ according to (\ref{numeric1}), which exhibits two possible solutions which lead to two different mass schemes. 
First, the {\it traditional scheme}, where $Y_t/Y_U \approx 135.05$ is required to fit the experimental masses, and where small variations of $Y_U$ imply large variations of $Y_t$. This ratio implies an asymmetry factor $\epsilon _{Ut} \approx -0.985$. Second, we obtain a "{\it natural scheme}" where  $Y_t/Y_U \approx 1.03$, which is consistent with a symmetry where degenerated up-type Yukawa couplings is favored, with $\epsilon _{Ut} \approx -0.015$. Fig. \ref{fig5} shows $Y_D$ as function of the ratio $\nu _{\sigma}/m_T$ according to (\ref{numeric3}), where we assume for simplicity that $y_1 \sim c_1 \sim Y_D$. We see that large values of $Y_D$ require small values of the VEV $\nu _{\sigma}$ in relation to the mass of the $T$- quark. For example, if $m_T \sim 1$ TeV, to obtain $Y_D \sim 0.01$ and $0.1$, we require that $ \nu _{\sigma } \sim 70 $ and $7$ GeV, respectively.  Finally, we find from (\ref{ds-ratio}) that the large splitting between $m_d$ and $m_s$ is consequence of the existence of non-degenerated heavy massive quarks according to (\ref{numeric2}). For example, if $M_{J^2} \sim 1$ TeV, then $M_{J^1} \sim 20$ TeV to obtain the observable $m_d/m_s$ ratio. Thus, the model predict beyond standard model particles that produce as consequence a large mass hierarchy observed among the SM quark families.   

\section{Conclusions}

Extensions with abelian non-universal $U(1)'$ symmetry are very well-motivated models which involves a wide number of theoretical aspects. In this work, by requiring non-universality in the left-handed quark sector, we proposed a new $G_{sm} \times U(1)_X$ gauge model with $Z'$ family non-universal gauge couplings, where two new gauge parameters arise: the $Z'-$ mass and the $U(1)_X$ coupling constant. By using the experimental limits on Drell-Yan cross section at CERN-LHC, we found constraints on these parameters at 95\% CL, obtaining upper bounds as shown in Fig. \ref{fig1}. Invariant-mass distributions for top-pair final states were generated using the Calchep package at $8$ TeV c.m. energy at LHC, obtaining $Z'$ resonances at $m_{Z'}=2, 2.5$ and $3$ TeV for two cases: i.) adjustable and ii.) fixed values of the coupling $g_X$. 

On the other hand, with the addition of three exotic quark singlets, one scalar doublet and two scalar singlets, we obtained a free-anomaly theory with Yukawa interactions invariant under a global $Z_2 \times U(1)_{T_3}$ symmetry which predict zero-texture mass matrices for the ordinary SM quarks with three non-physical massless quarks (u, d, s), and three massive quarks (c, b, t) at the electroweak scale (GeV). However, if mixing between SM quarks and the new exotic quarks consistent with the symmetries and the cancelation of the chiral anomalies are taken into account, the lightest quarks (u, d, s) obtain masses at the MeV scales. Predictable mass ratios were obtained with few free parameters, and hierarchical structures arose "naturally" without large tuning of the Yukawa couplings. These hierarchies can be understood as follows:

\begin{itemize}

\item[-] $m_{u,d,s}/m_{c,b,t}:$ Due to the non-universal $U(1)_X$ gauge symmetry and a global $Z_2$ symmetry, the Yukawa interactions among ordinary matter lead to zero-texture mass matrices of the up- and down-type quarks. However, since an additional $U(1)_{T_3}$ global symmetry is required only for the down sector, the mass structure between up- and down-type quarks is not equivalent: while in the up sector the c- and t-quarks acquire masses at the scale $\nu _1 \sim $ GeV, in the down sector only the b-quark obtains mass at this scale through $\nu _2$. Thus, the three quarks (u, d, s) remain massless. On the other hand, to cancel the chiral anomalies, three exotic quarks $T, J^{1,2}$ (and one neutrino singlet $(\nu _R)^c$) are introduced. The interactions between the massless and the exotic quarks, provide masses to the former. These masses are inverse in the heavy quark masses, obtaining a see saw-type mechanisms: $m_{u,d,s} \sim \nu \nu_{\sigma }/\nu _{\chi } \sim $ [GeV$^2$]$/[TeV]$ $\sim$ [MeV].
     

\item[-] $m_d/m_s:$ Since the d- and s-quarks acquire MeV scale masses through interactions with two non-degenerate heavy quarks $J^{1,2}$, the ratio $m_d/m_s$ is not one according to (\ref{ds-ratio}). If $m_{J^1}>m_{J^2}$ then $m_d/m_s<1$.

\item[-] $m_c/m_t:$ Although both c- and t-quarks acquire masses at the scale of GeV, they are non-degenerate due to the likeness between Yukawa constants in the natural scheme. Thus, a large ratio $m_t/m_c$ arises naturally if the top coupling $Y_t$ is slightly different to the other up-type quark coupling $Y_U$. Indeed, if the flavor symmetry between families 2 and 3 is exact in the up sector ($\epsilon _{Ut}=0$), the charm quark becomes massless and only the top quark acquire mass.      

\end{itemize}       

\section*{acknowledgment}

This work was supported by Colciencias

\newpage

\begin{table}[]
\begin{center}
\caption{\small Ordinary SM particle content, with $i=$1,2,3} \vspace{-0.5cm}
\label{tab:SM-espectro}
\begin{equation*}
\begin{tabular}{c c c c}
\hline\hline
$Spectrum$ & $G_{sm}$ & $U(1)_{X }$ & $Feature$\\ \hline \\
$\ 
\begin{tabular}{c}
$q^i_{L}=\left( 
\begin{array}{c}
U^i \\ 
D^i 
\end{array}%
\right) _{L}$ 
\end{tabular}%
\ $ & 
$(3,2,1/3)$
&
$\ 
\begin{tabular}{c}
$1/3$ for $i=1$ \\ 
$0$ for $i=2,3$%
\end{tabular}
\ $
&
chiral
\\ \\
$U_R^i$
& 
$(3^*,1,4/3)$
&
\begin{tabular}{c}
$2/3$ 
\end{tabular}
&
chiral
\\ \\
$D_R^i$
&
$(3^*,1,-2/3)$
&
\begin{tabular}{c}
$-1/3$ 
\end{tabular}
&
chiral
\\ \\
$\ 
\begin{tabular}{c}
$\ell ^i_{L}=\left( 
\begin{array}{c}
\nu ^i \\ 
e^i 
\end{array}%
\right) _{L}$ \\
\end{tabular}
\ $
& 
$(1,2,-1)$
&
$-1/3$ 
&
chiral
\\  \\
$e_R^i$
&
$(1,1,-2)$
&
\begin{tabular}{c}
$-1$ 
\end{tabular}
&
chiral
\\  \\
$\ 
\begin{tabular}{c}
$\phi _{1}=\left( 
\begin{array}{c}
\phi _1^{+} \\ 
\frac{1}{\sqrt{2}}(\nu _1+\xi _1+i\phi _1^0) 
\end{array}%
\right) $
\end{tabular}
\ $
&
$(1,2,1)$
&
$2/3$
&
Scalar Doublet
\\  \\
$\ 
\begin{tabular}{c}
$W _{\mu }=\left( 
\begin{array}{cc}
W _{\mu }^{3}&\sqrt{2}W_{\mu }^+ \\ 
\sqrt{2}W_{\mu }^-& -W _{\mu }^{3}
\end{array}%
\right) $
\end{tabular}
\ $
&
$(1, 2 \times 2^*, 0)$
&
$0$
&
Vector
\\  \\
$\ 
B_{\mu }
\ $
&
$(1, 1, 0)$
&
$0$
&
Vector
\end{tabular}%
\end{equation*}%
\end{center}
\end{table}

\begin{table}[tbp]
\begin{center}
\caption{\small Exotic non-SM particle content, with $n=$1,2}\vspace{-0.5cm}
\label{tab:exotic-espectro}
\begin{equation*}
\begin{tabular}{c c c c}
\hline\hline
$Spectrum$ & $G_{sm}$ & $U(1)_{X }$ & $Feature$\\ \hline \\
$\ 
\begin{tabular}{c}
$T_L$ 
\end{tabular}%
\ $ & 
$(3,1,4/3)$
&
$1/3$
&
quasi-chiral
\\ \\
\begin{tabular}{c}
$T_R$
\end{tabular}
& 
$(3^*,1,4/3)$
&
$2/3$
&
quasi-chiral
\\ \\
\begin{tabular}{c}
$J_L^n$ 
\end{tabular}
&
$(3,1,-2/3)$
&
$0$
&
quasi-chiral
\\ \\ 
$\ 
\begin{tabular}{c}
$J_R^n$  
\end{tabular}
\ $
& 
$(3^*,1,-2/3)$
&
$-1/3$ 
&
quasi-chiral
\\ \\
\begin{tabular}{c}
$(\nu _R^i)^c$ 
\end{tabular}
&
$(1,1,0)$
&
$-1/3$
&
Majorana
\\ \\ 
\begin{tabular}{c}
$N _R^i$ 
\end{tabular}
&
$(1,1,0)$
&
$0$
&
Majorana
\\ \\
$\ 
\begin{tabular}{c}
$\phi _{2}=\left( 
\begin{array}{c}
\phi _2^{+} \\ 
\frac{1}{\sqrt{2}}(\nu _2+\xi _2+i\phi _2^0) 
\end{array}%
\right) $ 
\end{tabular}
\ $
&
$(1,2,1)$
&
$1/3$
&
Scalar doublet
\\ \\
\begin{tabular}{c}
$\chi _0 = \frac{1}{\sqrt{2}}(\nu _{\chi}+\xi _{\chi }+i\zeta _{\chi})$ 
\end{tabular}
&
$(1,1,0)$
&
$-1/3$
&
Scalar singlet
\\ \\
\begin{tabular}{c}
$\sigma _0 = \frac{1}{\sqrt{2}}(\nu _{\sigma}+\xi _{\sigma}+i\zeta _{\sigma})$ 
\end{tabular}
&
$(1,1,0)$
&
$-1/3$
&
Scalar singlet
\\  \\ 
$B'_{\mu}$ 
&
$(1,1,0)$
&
$0$
&
Vector
\end{tabular}%
\end{equation*}%
\end{center}
\end{table}

\begin{table}[]
\begin{center}
\caption{\small Vector and Axial couplings for the weak neutral currents $Z$ (SM-type) and $Z'$ (non-SM type) and for each fermion, with $i=1,2,3$, $a=2,3$ and $n=1,2$} \vspace{-0.5cm}
\label{tab:vector-axial-couplings}
\begin{equation*}
\begin{tabular}{c c c c c c}
 \hline \hline \vspace{0.1cm} 
$Fermion$&$v_f^{SM}$&$a_f^{SM}$&&$v_i^{NSM}$&$a_i^{NSM}$   \\  \hline \vspace{0.2cm}
$\nu ^i$ &$1/2$&$1/2$&&$1/3$&$1/3$ \\ \vspace{0.2cm}
$(\nu ^i)^c$&$0$&$0$&&$1/3$&$1/3$ \\ \vspace{0.2cm}
$N^i$&$0$&$0$&&$0$&$0$ \\ \vspace{0.2cm}
$e^i$&$-1/2+2S_W^2$&$-1/2$&&$4/3$&$-2/3$ \\ \vspace{0.2cm}
$U^1$&$1/2-4S_W^2/3$&$1/2$&&$-1$&$1/3$\\ \vspace{0.2cm}
$U^a$&$1/2-4S_W^2/3$&$1/2$&&$-2/3$&$2/3$\\ \vspace{0.2cm}
$D^1$&$-1/2+2S_W^2/3$&$-1/2$&&$0$&$-2/3$\\ \vspace{0.2cm}
$D^a$&$-1/2+2S_W^2/3$&$-1/2$&&$1/3$&$-1/3$\\ \vspace{0.2cm}
$T$ &$-4S_W^2/3$&$0$&&$-1$&$1/3$\\ \vspace{0.2cm}
$J^n$&$2S_W^2/3$&$0$&&$1/3$&$-1/3$
\end{tabular}
\end{equation*}
\end{center}
\end{table}

\newpage


\end{document}